\documentclass[twocolumn,superscriptaddress,floatfix,prl]{revtex4}
\UseRawInputEncoding
\usepackage{amsmath,amsfonts,amssymb}
\usepackage{pdfpages}
\usepackage{graphicx}
\usepackage{xcolor}
\usepackage{xr-hyper}
\usepackage{hyperref}
\begin{document}
\title{Loss-driven topological transitions in lasing}
\author{Grazia Salerno}
\email{gresalerno+physics@gmail.com}
\author{Rebecca Heilmann}
\author{Kristian Arjas}
\author{Kerttu Aronen}
\author{Jani-Petri Martikainen}
\author{P\"{a}ivi T\"{o}rm\"{a}}
\email{paivi.torma@aalto.fi}
\affiliation{Department of Applied Physics, Aalto University School of Science, P.O. Box 15100, Aalto, FI-00076, Finland}
\begin{abstract}
We experimentally observe lasing in a hexamer plasmonic lattice and find that when tuning the scale of the unit cell, the polarization properties of the emission change. By a theoretical analysis we identify the lasing modes as quasi bound states in continuum (quasi-BICs) of topological charges of zero, one or two. A T-matrix simulation of the structure reveals that the mode quality(Q)-factors depend on the scale of the unit cell, with highest-Q modes favored by lasing. The system thus shows a loss-driven transition between lasing in modes of trivial and high-order topological charge. 
\end{abstract}
\maketitle

Bound states in the continuum (BICs) are peculiar eigenstates of quantum or classical wave systems: their energies lie within a continuum of other states but they are nevertheless completely uncoupled from those states. For optical structures such as gratings, cavities and photonic crystals this means that their BIC modes are decoupled from free-space radiation and appear as dark states: light cannot in- or out-couple from/to the far-field at the energy and momentum of the BIC~\cite{SoljacicRev, hwang2022reviewBIC}.  
From the fundamental perspective, photonic BICs can be viewed as vortex centers of light polarization. When one measures the polarization of light radiated by the structure from different angles, corresponding to different momentum, the polarization winds around the BIC momentum. Due to this winding, it is impossible to define what would be the polarization exactly at the BIC momentum, which therefore must appear dark, similarly as the core of a vortex is empty. This description has naturally connected the BICs to the realm of more general topological phenomena, including topological robustness and topological transitions~\cite{OzawaRev, RiderRev_Perspective, KivsharRev, RiderRev_Advances}. 
The polarization vortex is associated with the existence of a protected and quantized topological charge which tells how many times the polarization winds in a path circulating the BIC momentum~\cite{Alu, SoljacicTopoBic}. This charge is conserved and cannot be removed with small perturbations. In order to observe BICs, they can be weakly coupled to the radiative continuum via an intentionally designed leakage mechanism or due to material loss~\cite{LasingAzzam, KodigalaLasingBIC,LasingHa}; the presence of an edge can also act as a leakage channel~\cite{heilmann2022,Asamoah2022}. Such leaky or lossy BICs are called quasi-BICs. Quasi-BICs can have extremely large Q-factors~\cite{iwahashi2011higher, zito2019observation, wang2020generating, huang2020ultrafast, dyakov2021photonic, zhang2021bound, hwang2021ultralow, wu2022active, kang2022high}, which is promising for numerous applications in lasing, filtering, or sensing~\cite{GentryDarkStateLasers, KodigalaLasingBIC, LasingHa, guan2020engineering, guan2020quantum, silva2020multiple}. 

BICs of topological charge $|q|=1$ have been observed in many systems, from photonic crystals to nanoplasmonic structures~\cite{Alu, SoljacicTopoBic, de2019index, soljacic_transition_sameq, heilmann2022, Hakala, Asamoah2022, KodigalaLasingBIC}.  Switching between non-degenerate BICs with different topological charge of $|q|\leq 1$ was recently achieved~\cite{Hakala}.
The interest in BICs with large topological charges ($|q|>1$) and their generation, evolution, or annihilation is only just starting. 
BICs of high-order charge are predicted to exist in systems with a six-fold rotational symmetry~\cite{SoljacicTopoBic, bai2021terahertz} and it has been theoretically proposed that a BIC of charge $|q|=2$ can undergo a topological transition and break into two BICs of charge $|q|=1$~\cite{yoda2020generation}. 
The first experimental studies demonstrating vortex beams generated by BIC of higher topological charge in photonic crystals have recently emerged~\cite{zhang2018observation, wang2020generating}, while observation of lasing in BICs of higher topological charges still lacks in plasmonic lattices.

In this Letter, we observe lasing in a plasmonic lattice from a BIC of topological charge $q=-2$ and demonstrate several topological transitions driven by losses as a structural parameter of the unit cell is changed. We use plasmonic nanoparticle arrays due to their strong dipole moments in the nanoparticles~\cite{QuantumPlasmonics, Wang, wang_structural_2018, VakevainenStrongCoupling, HakalaBEC, kravets_plasmonic_2018}.
The lasing mode is identified by using theory of coupled dipoles together with group theory and T-matrix calculations~\cite{Necada, NecadaQPMS}. We show that the energy of the modes as well as their loss properties change as the structural parameter is varied. The Q-factor of the modes changes, and the mode with the highest Q-factor lases. By comparing the theory with the experimental measurements, we characterize the polarization vorticity of the lasing mode and confirm topological transitions between a bright mode and quasi-BIC modes of topological charges $q=-2$, $q=-1$ and $q=+1$.

\begin{figure*}
    \centering
    \includegraphics[width=0.95\textwidth]{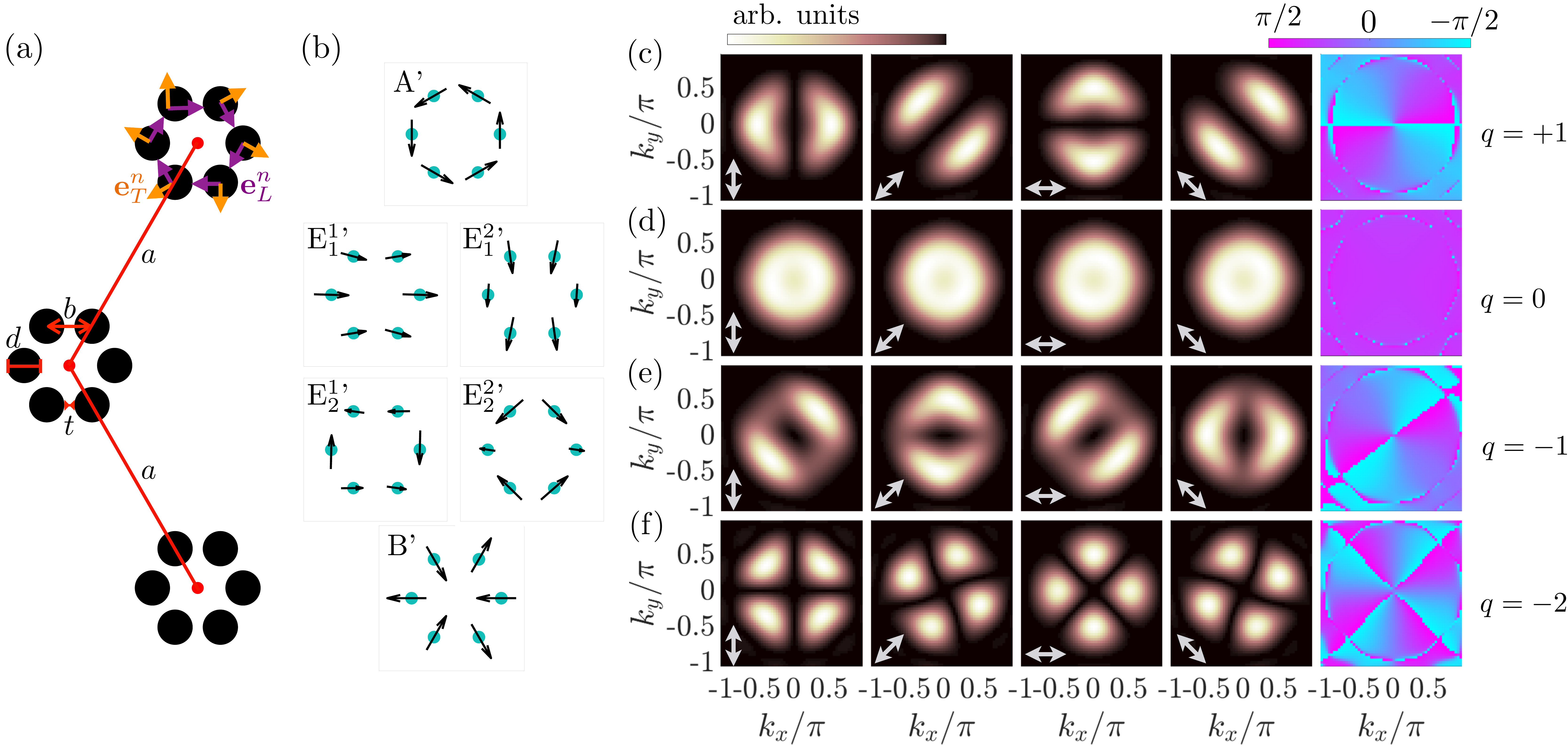}
    \caption{(a) The hexamer is composed of six particles (black disks) arranged in a regular hexagon of side length $b$. Purple and orange arrows represent the longitudinal $\mathbf{e}_L^n$ and transverse $\mathbf{e}_T^n$ unit vectors used in the simple hexamer model, having indicated only nearest-neighbor. The hexamer constitutes the unit cell of a triangular lattice of periodicity $a$ (red points). Due to a finite size of the particles, whose diameter is $d$, the edge-to-edge distance between nanoparticles is $t = b - d$.
    (b) Modes of an isolated hexamer, where the arrows correspond to the electric dipole moment. The modes are labeled according to the corresponding IRs of the symmetry group $C_{6}$.
    (c-f) Amplitude and phase of the four IRs, filtered with a polarizer oriented according to the arrow in each panel, while the phase in the last column is unfiltered. The modes and their respective topological charges are (c) $A'$: $q=+1$; (d) $E_1'$: $q=0$; (e) $E_2'$: $q=-1$, and (f) $B'$: $q=-2$. (d-e) show balanced superpositions of the degenerate modes with zero relative phase.}
    \label{fig:Fig1}
\end{figure*}
 
The topological charge of a BIC is defined as~\cite{SoljacicTopoBic}
\begin{equation}
    q = \frac{1}{2\pi} \oint_\mathcal{C} d \mathbf{k} \cdot \nabla_\mathbf{k} \Phi(\mathbf{k}),
    \label{topocharge_def}
\end{equation} 
where $\Phi(\mathbf{k}) = \arg[\mathbf{p}(\mathbf{k}) \cdot \hat{x} + i \mathbf{p}(\mathbf{k}) \cdot \hat{y}]$ is the phase of the polarization vector projected onto the $xy$ plane in momentum space
$\mathbf{p}(\mathbf{k}) = (\hat{x} \cdot \langle \mathbf{u}_\mathbf{k}(\mathbf{r}, z)\rangle)\hat{x} + (\hat{y} \cdot \langle \mathbf{u}_\mathbf{k}(\mathbf{r}, z) \rangle)\hat{y}$, and $\langle \mathbf{u}_\mathbf{k}\rangle $ indicates the spatial average of the periodic part of the electric field $\mathbf{E}_\mathbf{k}(\mathbf{r},z) = e^{i \mathbf{k}\cdot\mathbf{r}} \mathbf{u}_\mathbf{k}(\mathbf{r},z)$ at constant vertical distance $z$ from the lattice plane much larger than the unit cell size.
The charge $q$ counts the number of times the polarization vector $\mathbf{p}(\mathbf{k})$ goes around a specific $\mathbf{k}$ point, e.g.~the $\Gamma$-point. The sign of the topological charge gives the orientation of the winding.

We start by theoretically analyzing the modes of an isolated hexamer, Fig.~\ref{fig:Fig1}(a).
Each particle has an in-plane dipole moment $\mathbf{p}_i = (p_i^x, p_i^y)$, coupled to all other dipoles in a polarization-dependent way, as in Ref.~\cite{heilmann2022}. We set $\Omega_L$ ($\Omega_T$) to be the coupling between dipoles oriented longitudinally (transversely) to the link $\mathbf{e}_L^j= (\mathbf{r}_i - \mathbf{r}_{i+j})/|\mathbf{r}_i - \mathbf{r}_{i+j}|$ (such that $\mathbf{e}_T^j \cdot \mathbf{e}_L^j =0$) connecting particles $i$ and $i+j$ (typically for dipoles $\Omega_T\gg \Omega_L$). The bare dipole oscillation frequency is $\omega_0 \gg \Omega_{L,T}$.
The modes are found from
\begin{equation*}
    \ddot{\mathbf{p}}_i \!=\! \omega_0 \mathbf{p}_i + \! \sum_{j\neq i} \left[ \frac{\Omega_L}{R_j^3} (\mathbf{p}_{i+j} \!\cdot\! \mathbf{e}_L^j )\mathbf{e}_L^j \!+\! \frac{\Omega_T}{R_j^3} (\mathbf{p}_{i+j} \!\cdot \! \mathbf{e}_T^j )\mathbf{e}_T^j \right] ,
\end{equation*}
where $R_j = |\mathbf{r}_i - \mathbf{r}_{i+j}|$. In the basis $(p_i^x, p_i^y)$, we find twelve modes centered around $\omega_0$. The spatial structure of the lower energy set of eigenmodes is shown in Fig.~\ref{fig:Fig1}(b) where arrows correspond to the electric dipole moment orientation. The modes belong to the symmetry class of the $C_{6}$ group; the singlets to the $A$ and $B$ irreducible representations (IRs), the doublets to the $E_1=(E_1^1, E_1^2)$ and the $E_2=(E_2^1, E_2^2)$ IRs. The ordering of these modes does not depend on the interparticle distance, in the isolated hexamer case as losses are not included.
From Finite-Element-Method simulations on arrays with periodic boundary conditions, we find the same modes, and also another set of six modes ~\cite{SuppMat} that have a larger out-of-plane character. Those are not excited in our setup and therefore we will focus on the $A'$, $E_1'$, $E_2'$ and $B'$ modes.

From the electric dipole moment orientation in each mode, we calculate the real-space diffraction pattern generated by the nanoparticle array, together with the far field in momentum space. 
BICs have distinct momentum-space features, as visible in Fig.~\ref{fig:Fig1}(c-f), where the far field is obtained by Fourier transforming the real space field with a polarization filter oriented according to the arrow in each panel; these calculations are done as in Refs.~\cite{GuoLasing, heilmann2022}.
The last column in Fig.~\ref{fig:Fig1}(c-f) shows the phase of each mode's polarization calculated from the Stokes parameters ($\Phi = \frac 1 2 \arctan(S_1/S_2)$) without any polarization filter.
Fig.~\ref{fig:Fig1}(d) shows that the $E_1'$ doublet modes are bright modes, since their emission in momentum space is featureless along all polarization directions while the phase is uniform. These modes have no topological charge ($q=0$), and correspond to dipoles oscillating in phase along the same direction.
All other modes are quasi-BICs, and the dipoles oscillate in such a way that the emission from $\mathbf{k}=0$ is dark. With a polarization filter, we see a non-zero emission in the vicinity of the $\Gamma$-point, on two or four opposite lobes of a donut-shaped amplitude pattern~\cite{iwahashi2011higher, LasingWu, hwang2021ultralow, Hakala}. As the polarization filter is rotated from vertical polarization in a clockwise direction, the lobes rotate as well, following the non-trivial phase polarization (see last column).
In particular, the polarized amplitude of mode $A'$ in Fig.~\ref{fig:Fig1}(c) has two lobes that rotate in the clockwise direction, which correspond to $q=1$. Accordingly, the phase in the last column of Fig.~\ref{fig:Fig1}(c) shows a $2\pi$ rotation, corresponding to a unit winding. Similarly, the doublet modes $E_2'$ have $q=-1$ in Fig.~\ref{fig:Fig1}(e), since the two lobes rotate counter-clockwise; this is also visible in the different phase winding. Finally, the charge $q=-2$ of mode $B'$ in Fig.~\ref{fig:Fig1}(f) is evidenced by a four-lobe rotation in a counter-clockwise direction. Consequently, the phase in the last column has an overall $4\pi$ winding.
The topological charge $q_\text{IR}$ of a mode corresponding to an IR can be derived~\cite{SuppMat} from the n-fold rotational symmetry $C_n$:
\begin{equation}
    q_{\text{IR}} =  1 - \frac{n}{2\pi}\arg(\epsilon_\text{IR}),
    \label{topocharge}
\end{equation}
where $\epsilon_\text{IR}$ can be found in character tables. Values given by Eq.~(\ref{topocharge}) for $C_6$ equal those given by our analysis of coupled dipoles in Fig.~\ref{fig:Fig1}. 

\begin{figure*}
\includegraphics[width=0.95\textwidth]{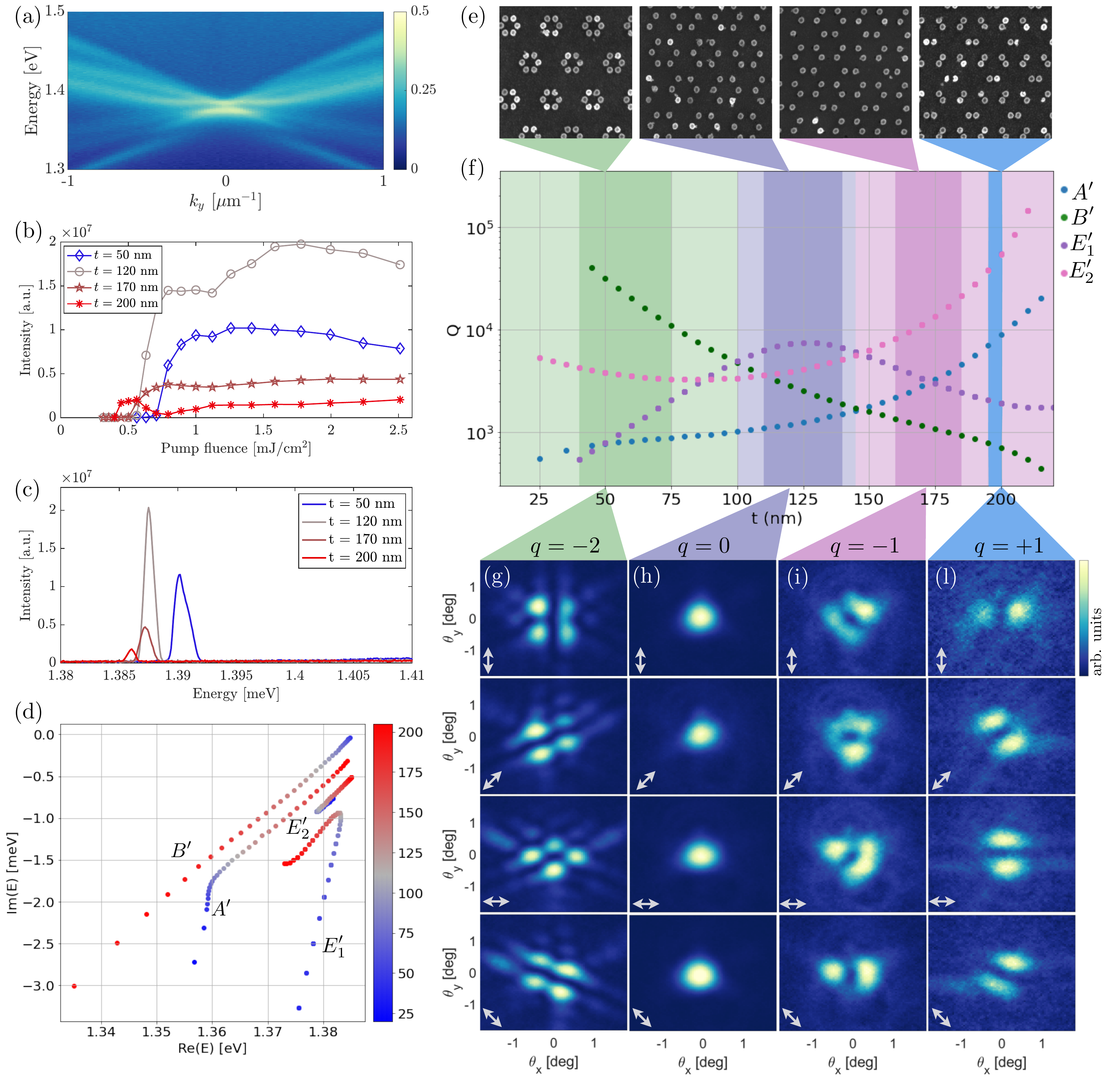}
\caption{(a) Energy dispersion of an array of cylindrical gold nanoparticles of diameter $d=80$~nm, height $h=50$~nm, arranged in a triangular array of periodicity $a = 680$~nm, with $t = 50$~nm, as a function of the $y$-component of the in-plane momentum. (b)~Peak intensity as a function of the pump fluence.
(c)~Spectra of the lasing emission at a pump fluence of 1.78~mJ/cm$\mathrm{^{2}}$.
(d)~Energies of the $\Gamma$-point modes in the complex plane calculated with the T-matrix approach, vertically shifted for clarity (without the shift in the Supplemental Material~\cite{SuppMat}). The color scale indicates the value of $t$ [nm]. 
(e)~SEM images of arrays with $t=$ 50, 120, 170, 200~nm from left to right.
(f)~Q-factor of the modes as a function of $t$ [nm]. The light-colored region indicates the highest Q-factor mode, which is expected to be the lasing mode. The corresponding experimental observations are marked with the matching dark-colored regions.
(g-l) Angle-resolved images of the lasing emission with different polarization filters in the detection path, indicated by the double-headed arrow in the lower left corner of each image. The array is pumped with LCP polarization and pump fluences of 1~ mJ/cm$\mathrm{^{2}}$ for t = 50 nm (g), 1.58~ mJ/cm$\mathrm{^{2}}$ for t = 120 (h) and t= 170 nm (i), and 0.63~ mJ/cm$\mathrm{^{2}}$ for t= 200 nm (l).}
\label{fig:Fig2}
\end{figure*}    
 
We now consider an extended array composed of hexamers arranged in a finite-size triangular lattice.
The edge-to-edge distance between nanoparticles of the same hexamer is $t$. The geometry of the system is depicted in Fig.~\ref{fig:Fig1}(a). 
We fabricated arrays of cylindrical gold nanoparticles with an edge-to-edge distance that is varied between $45<t<200$~nm. The system hosts dispersive plasmonic-photonic modes~\cite{QuantumPlasmonics, Wang, wang_structural_2018,  kravets_plasmonic_2018}. The dispersion of such an array for $t = 50$~nm is shown in Fig.~\ref{fig:Fig2}(a). The arrays were combined with a solution of fluorescent dye molecules (IR-140) with a concentration of 10 mM. Under optical pumping with a left-circularly polarized (LCP) 100 fs laser pulse (center wavelength $800$~nm (1.55 eV)), we observe a non-linear increase in the emission at $k_y=0$ with increasing pump fluence as shown in Fig.~\ref{fig:Fig2}(b): arrays with different $t$ show different behavior. 
Furthermore, the peak intensity differ as well: the lasing peak of the $t = 120$~nm  array is about one order of magnitude larger than the one for the $t = 200$~nm array. More details of the experiments are given in the Supplemental Material~\cite{SuppMat}, where we also show the real space emission from the arrays in the lasing regime.

The different thresholds, peak intensities, and lasing energies in Fig.~\ref{fig:Fig2}(b-c) hint towards having a qualitatively different mode lasing for different array geometries. To understand this, we recorded angle- and polarization-resolved measurements of the lasing emission from arrays with a wide range of different hexamer sizes.
In Fig.~\ref{fig:Fig2}(g-l) we show angle-resolved images of arrays with $t =$ 50, 120, 170, and 200~nm with different polarization filters.
For $t = 50$~nm, we see that the momentum-space emission in Fig.~\ref{fig:Fig2}(g) mainly concentrates in four lobes which rotate in a counter-clockwise direction, compatibly with the $B'$ mode of $q=-2$. For $t = 120$~nm in Fig.~\ref{fig:Fig2}(h), there is only a single lobe with no visible change as the polarization filter is rotated; this mode is the doublet mode $E_1'$, which has no topological charge.
As we further increase the size of the unit cell, for $t = 170$~nm, in Fig.~\ref{fig:Fig2}(i) we see two lobes rotating in a counter-clockwise direction, compatibly with the doublet $E_2'$ of $q=-1$.
For even higher $t = 200$~nm in Fig.~\ref{fig:Fig2}(l), the momentum-space emission comes from two lobes that rotate in a clockwise direction, as for mode $A'$ with $q=+1$. 
These observations directly correspond to the amplitudes and phases theoretically calculated in Fig.~\ref{fig:Fig1}(c). 

We have employed T-matrix calculations~\cite{Necada, NecadaQPMS} (see also~\cite{SuppMat}) to explain why there are four regimes in which the lasing mode corresponds to a different IR. We obtain the energy of the modes at the $\Gamma$-point by varying $t$ in the experimentally relevant parameter range, and plot their real and imaginary parts in the complex plane, in Fig.~\ref{fig:Fig2}(d). We see that the mode energies mostly lie on a linear trajectory. The imaginary part of the energy accounts for the losses, which in our system are radiative and/or Ohmic losses.
The modes $A'$, $B'$, and $E_2'$ are dark modes that do not radiate to the continuum: for this reason, they only experience Ohmic losses, and the complex energy curve has the same slope for all three. The bright mode $E_1'$ is subjected to both Ohmic and radiative losses, thus its energy slope for $t<150$~nm is steeper than for the three dark modes. However, for $t>150$~nm the slope of $E_1'$ gets smaller, indicating that the mode experiences less radiative losses. In general, all modes' losses behave differently for small ($t<100$~nm) and large ($t>210$~nm) hexamers; for example, at small $t$, mode $B'$ seems to have a very small imaginary part, and the same is true for mode $A'$ for large $t$ instead. As the modes have different radiative properties when the hexamer size $t$ is varied, a loss-induced transition between the modes could occur in lasing (see \cite{SuppMat} for results also on a quadrumer array). The change of losses for the various modes with different $t$ is supported by FEM simulations where we see that the modes' electric field intensity distribution in the unit cell changes significantly with changing $t$, despite the symmetry of the modes being the same~\cite{SuppMat}. The field hot-spots away from the nanoparticles can contribute to how lossy a given mode and hence its quality factor Q, and also its coupling to the gain medium.

From the complex energies of the modes in Fig.~\ref{fig:Fig2}(d), we calculated the Q-factor of each IR as a function of the hexamer size, see Fig.~\ref{fig:Fig2}(f). 
For small hexamers (light green region $t<100$~nm) the singlet $B'$ with $q=-2$ has the largest Q-factor; the experiment shows lasing from a quasi-BIC of charge $q=-2$ in the dark green region.
For intermediate hexamers (light violet region for $100<t<145$~nm) the largest Q-factor mode is the $E_1'$ doublet with $q=0$, which in the experiment lases for the dark violet region. For large hexamers (light pink region $t>150$~nm) the other doublet $E_2'$ with $q=-1$ has the largest Q-factor in T-matrix simulations and the experiment evidences lasing with this topological charge in the dark pink region. Lastly, the experiment shows lasing in the $A'$ IR with $q=+1$ for $t\approx 200$~nm, although this mode has a lower Q-factor than $E_2'$. We attribute this inconsistency to the different spatial overlap of the mode with the gain medium mentioned above, causing mode competition between $A'$ and $E_2'$~\cite{SuppMat}. 

In an equilibrium topological transition, the topological character of the ground state changes. In our non-equilibrium lasing system, modes of different topological charge coexist with nearly identical energies, but the mutual ordering of their Q-factors changes with a structural parameter. Therefore lower losses (effectively higher gain) can drive a transition of lasing between modes of different topological charge. We have demonstrated lasing in a topologically trivial bright mode and quasi-BIC states of topological charges $q=\pm1$ and $q=2$. Since topological charges take only discrete values, there must be a transition, not a smooth crossover between the regimes. Our results inspire future studies of the transition boundaries: is there multimode lasing, no lasing, bistability, fractional topological charges? Furthermore, the edge states between areas of different structural parameter may have properties distinct from the edge states of equilibrium systems. Other possible outlooks include the study of splitting the doubly-degenerate BICs at high-symmetry points of the Brillouin zone and the fate of their topological charge when degeneracy is removed. In general, our results introduce a versatile new platform for studies of non-Hermitian, out-of-equilibrium topological physics~\cite{yu2016loss, UedaRev, BergholtzRev}.

\begin{acknowledgments}
We thank Javier Cuerda for useful discussions on the finite element method simulations. This work was supported by the Academy of Finland under project number 318937 (PROFI), the Academy of Finland Flagship Programme, Photonics Research and Innovation (PREIN), project numbers 320167 and by Centre for Quantum Engineering (CQE) at Aalto University. Part of the research was performed at the OtaNano Nanofab cleanroom (Micronova Nanofabrication Centre), supported by Aalto University. We acknowledge the computational resources provided by the Aalto Science-IT project. GS has received funding from the European Union's Horizon 2020 research and innovation programme under the Marie Sk\l{}odowska-Curie grant agreement No 101025211 (TEBLA).
\end{acknowledgments}



\section*{Supplementary material}

# Loss-driven topological transitions in lasing

Grazia Salerno,[1, *] Rebecca Heilmann,[1] Kristian Arjas,[1] Kerttu Aronen,[1] Jani-Petri Martikainen,[1] and Päivi Törmä[1, †]

[1]*Department of Applied Physics, Aalto University School of Science, P.O. Box 15100, Aalto, FI-00076, Finland*

## Supplemental Material

### TOPOLOGICAL CHARGES FROM ROTATIONAL SYMMETRY AT THE Γ-POINT

We show that the $n$-fold rotational symmetry $C_n$ of a system can be applied to determine the topological charge of its modes, which are expressed as irreducible representation of $C_n$. From the definition in Eq. (2) of the main text, we choose the contour $\mathcal{C}$ to be a circular path of radius $k = |\mathbf{k}|$ centered around the origin, and evaluate the integral in polar coordinates. The gradient in Eq. (2) in the main text becomes

$$\nabla_{\mathbf{k}}\Phi = \frac{\partial}{\partial k}\Phi \hat{k} + \frac{1}{k}\frac{\partial}{\partial \varphi}\Phi \hat{\varphi}. \tag{S1}$$

Along the path, the line element becomes $d\mathbf{k} \to k d\hat{\varphi}$ and thus Eq. (2) is simplified to

$$q = \frac{1}{2\pi}\oint \frac{1}{k}\frac{\partial}{\partial \varphi}\Phi \hat{\varphi} \cdot k d\hat{\varphi} = \frac{1}{2\pi}\oint \frac{\partial \Phi}{\partial \varphi}d\varphi = \frac{1}{2\pi}\oint d\Phi. \tag{S2}$$

Using the rotational symmetry $C_n$, the path can be broken into $n$ identical parts

$$q = \frac{n}{2\pi}\int_{\Phi(\varphi)}^{\Phi(\varphi+2\pi/n)} d\Phi(\varphi) = \frac{n}{2\pi}[\Phi(2\pi/n + \varphi)) - \Phi(\varphi)]. \tag{S3}$$

The phase $\Phi(2\pi/n+\varphi)$ can then be related to the phase $\Phi(\varphi)$ upon rotating the system. A rotation by $2\pi/n$ modifies the field $E$ of a mode of a certain IR in the following way:

$$\mathcal{R}_{2\pi/n}(E_x(\varphi) + iE_y(\varphi)) = \mathcal{R}_{2\pi/n}(E_0(\varphi)e^{i\varphi}) = E_0(\varphi + \frac{2\pi}{n})e^{i(\varphi+2\pi/n)} = \epsilon_{\text{IR}}^* E_0(\varphi)e^{i2\pi/n}e^{i\varphi}.$$

Therefore, the phase of the field transforms as

$$\Rightarrow \Phi(\varphi + 2\pi/n) = \arg(\epsilon_{\text{IR}}^* e^{i2\pi/n}E_0(\varphi)e^{i\varphi}) = \arg(\epsilon_{\text{IR}}^*) + 2\pi/n + \Phi(\varphi) + m \cdot 2\pi.$$

The first term comes from the rotational symmetry of the group and it is different for different IRs of the system. Tabulated values of $\epsilon_{\text{IR}}^*$ can be found in character tables [S1]. Note that we are using the conjugate of $\epsilon_{\text{IR}}$, this is because character tables are defined for clock-wise rotations while our integral is counter-clockwise. The second term is from the definition of our field orientation. The third term is the phase at $\varphi : \Phi(\varphi) = \arg[E(\varphi)e^{i\varphi}]$. The fourth term simply states that there is a $2\pi$ ambiguity when taking the phase. With this connection of $\Phi(2\pi/n+\varphi)$ to $\Phi(\varphi)$, one can express the topological charge of each IR (with $m = 0$ for simplicity, and using $\arg(\epsilon_{\text{IR}}^*) = -\arg(\epsilon_{\text{IR}})$)

$$q_{\text{IR}} = \frac{n}{2\pi}\left(\frac{2\pi}{n} - \arg(\epsilon_{\text{IR}})\right) = 1 - \frac{n}{2\pi}\arg(\epsilon_{\text{IR}}). \tag{S4}$$

In Table I we summarize the topological charges of the modes belonging to different irreducible representation with various rotational symmetries $C_n$. We see that higher topological charges $|q| > 1$ are possible for unit cells that obey $C_6$ (hexamers) or $C_8$ (octamers), and generalizations for $C_n$ with $n > 8$ are straightforward. Table I is consistent with the results given in [S2] for the allowed topological charges at Γ-point for singly degenerate modes, but it is extended to include also the doubly degenerate modes. Notably, these doublets can also be BICs with nontrivial charges $|q| \geq 1$, and a further research direction is to study the splitting of these doubly-degenerate BICs and the fate of their topological charge.

| Symmetry | IR | $\arg(\epsilon_{\text{IR}})$ | q |
|---|---|---|---|
| $C_2$ | A | 0 | +1 |
|  | B | $\pi$ | 0 |
| $C_3$ | A | 0 | +1 |
|  | $E^*$ | $2\pi/3$ | 0 |
| $C_4$ | A | 0 | +1 |
|  | B | $\pi$ | $-1$ |
|  | $E^*$ | $\pi/2$ | 0 |
| $C_5$ | A | 0 | +1 |
|  | $E_1^*$ | $2\pi/5$ | 0 |
|  | $E_2^*$ | $4\pi/5$ | $-1$ |
| $C_6$ | A | 0 | +1 |
|  | B | $\pi$ | $-2$ |
|  | $E_1^*$ | $\pi/3$ | 0 |
|  | $E_2^*$ | $2\pi/3$ | $-1$ |
| $C_8$ | A | 0 | +1 |
|  | B | $\pi$ | $-3$ |
|  | $E_1^*$ | $\pi/4$ | 0 |
|  | $E_2^*$ | $\pi/2$ | $-1$ |
|  | $E_3^*$ | $3\pi/4$ | $-2$ |

TABLE I. Topological charges of the irreducible representation (IRs) for point group $C_n$ predicted by using Eq. (S4). Doubly degenerate IRs are indicated with an asterisk (*).

## T-MATRIX SIMULATIONS OF THE EIGENMODES

We provide a brief explanation of the T-Matrix method in the context of plasmonic lattices. For full derivation and access to the code, please refer to [S3, S4].

Light scattering from a spherical boundary embedded in linear and isotropic medium follows the Helmholtz-equation $(\nabla^2 - \kappa^2)\mathbf{E} = 0$. In spherical coordinates, this equation has two solutions: the *regular* and *outgoing* vector spherical wave-functions. These are related to the incoming ($a$) and scattered ($f$) electric fields respectively. From the assumed linearity of the medium, there exists a linear relation between these two:

$$f = Ta, \tag{S5}$$

where $T$ is the T-matrix which fully describes the scattering properties of our spherical boundary. Note that the scatterer does not have to be a sphere, but must be enclosed inside the boundary. These T-Matrices can be solved analytically for spheres and numerically for other shapes.

In a lattice, there are multiple scatterers. The field felt by particle $p$ consists of the incident field $a_0$ and the scattered field from all other particles

$$a_p = a_0 + \sum_{q \neq p} W_{p \leftarrow q} f_q = a_0 + \sum_{q \neq p} W_{p \leftarrow q} T_q a_q, \tag{S6}$$

where $W_{p \leftarrow q}$ is a translation operator that moves the scattered field from $q$ to $p$. The equation can be written on an unit-cell level as

$$a = a_0 + WTa \Rightarrow (I - WT)a = a_0, \tag{S7}$$

where $a = [a_1, a_2, ...]^T$ contains the field coefficients of all of the particles in the unit cell and $W, T$ are the corresponding block-matrices. Expressing this in the terms of scattered fields, we find the mode-equation

$$(I - TW)f = f_0. \tag{S8}$$

The eigenmodes of the system are found as the modes that propagate inside the system, i.e. by setting the external fields $f_0 = 0$, and then looking for non-trivial solutions to the equation (S8).

In the simulations presented in this study, the T-matrices were generated from the Null-field method for cylinders. For the scatterer, the electrical properties were obtained from the Lorentz-Drude model of gold and for the background medium, a constant refractive index $n = 1.52$ was used. The system size in the simulations was infinite.

## FINITE ELEMENT SIMULATIONS OF THE EIGENMODES

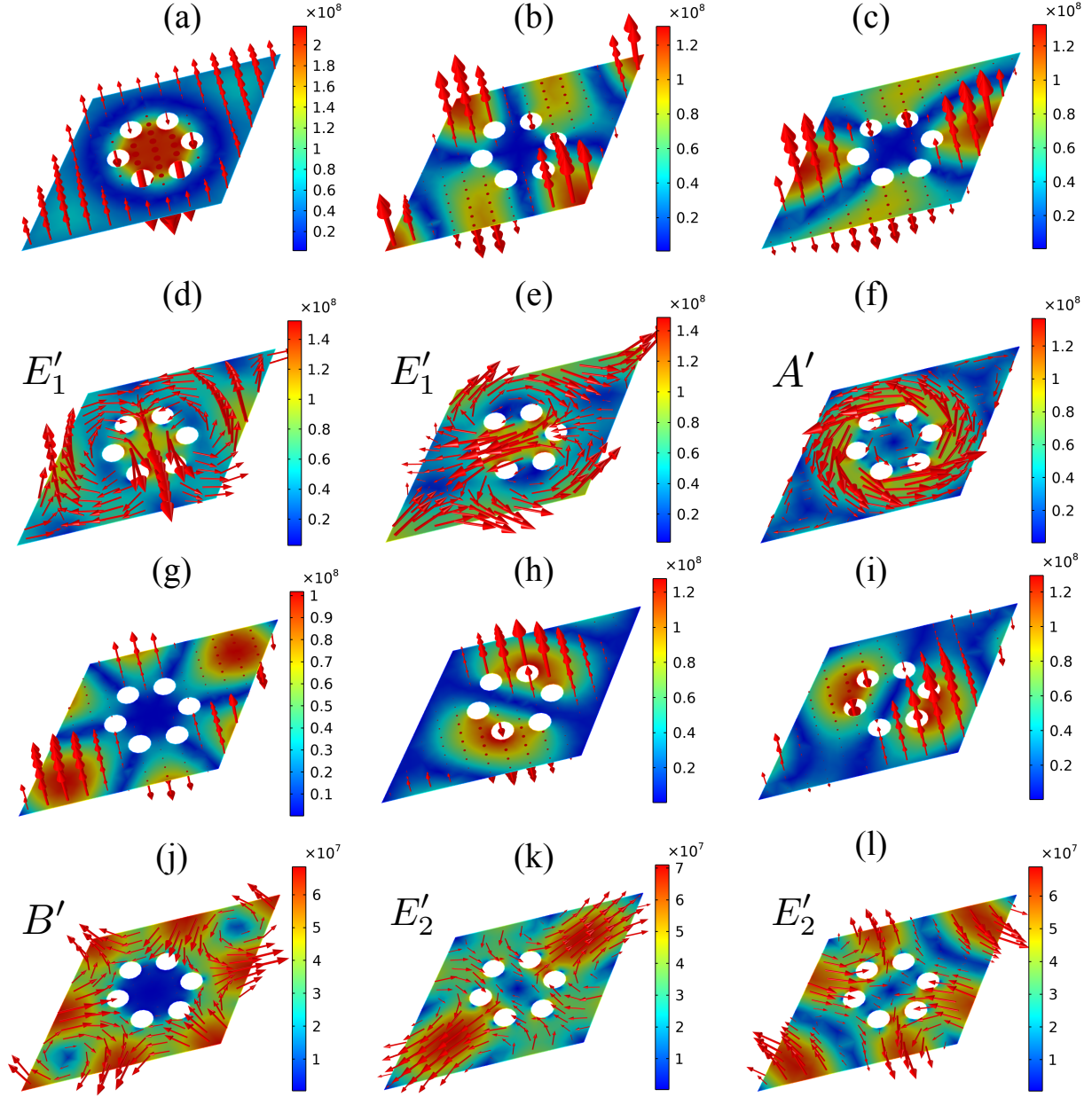

FIG. S1. Electric field magnitudes and directions for the eigenmodes in the array plane for the triangular lattice of hexamers. Results were computed with finite element method (COMSOL software) assuming 50 nm edge to edge distance for the cylindrical gold particles with a diameter of 80 nm and height of 50 nm. Background index of refraction was $n = 1.52$ and the triangular lattice had a periodicity of 680 nm. (b) and (c), (d) and (e), (h) and (i), and (k) and (l) are degenerate while remaining 4 modes ((a), (f),(g), and (j)) are singlets. (a-c) and (g-i) are out-of-plane modes, (d-f), (j-l) are the in-plane ones. The correspondence between the modes that we observed lasing is: (d-e) $E'_1$, (f) $A'$, (j) $B'$, (k-l) $E'_2$, which is also indicated in the figures.

4

We confirmed the validity of our theoretical approach by also completing finite element method simulations of our system using COMSOL to determine the eigenmodes of triangular lattice of hexamers. We found that the structures of the modes agree with the one found by our coupled dipole analysis and the T-matrix calculations. In Fig. S1 we show the electric fields in the array plane for all 12 modes when the edge to edge distance between particles was 50 nm. As is clear, 6 modes have out-of-plane character while 6 have in-plane character.

We show examples of the eigenmodes electric fields in the array plane for two different hexamer sizes in Fig. S2. Here figures only show the modes where the electric field is predominantly in the array plane as they are the ones playing a role in our experiments. There are also 6 modes with out of plane character which are not shown. Those are shown in Fig. S1 for the case with 50 nm edge to edge distance.

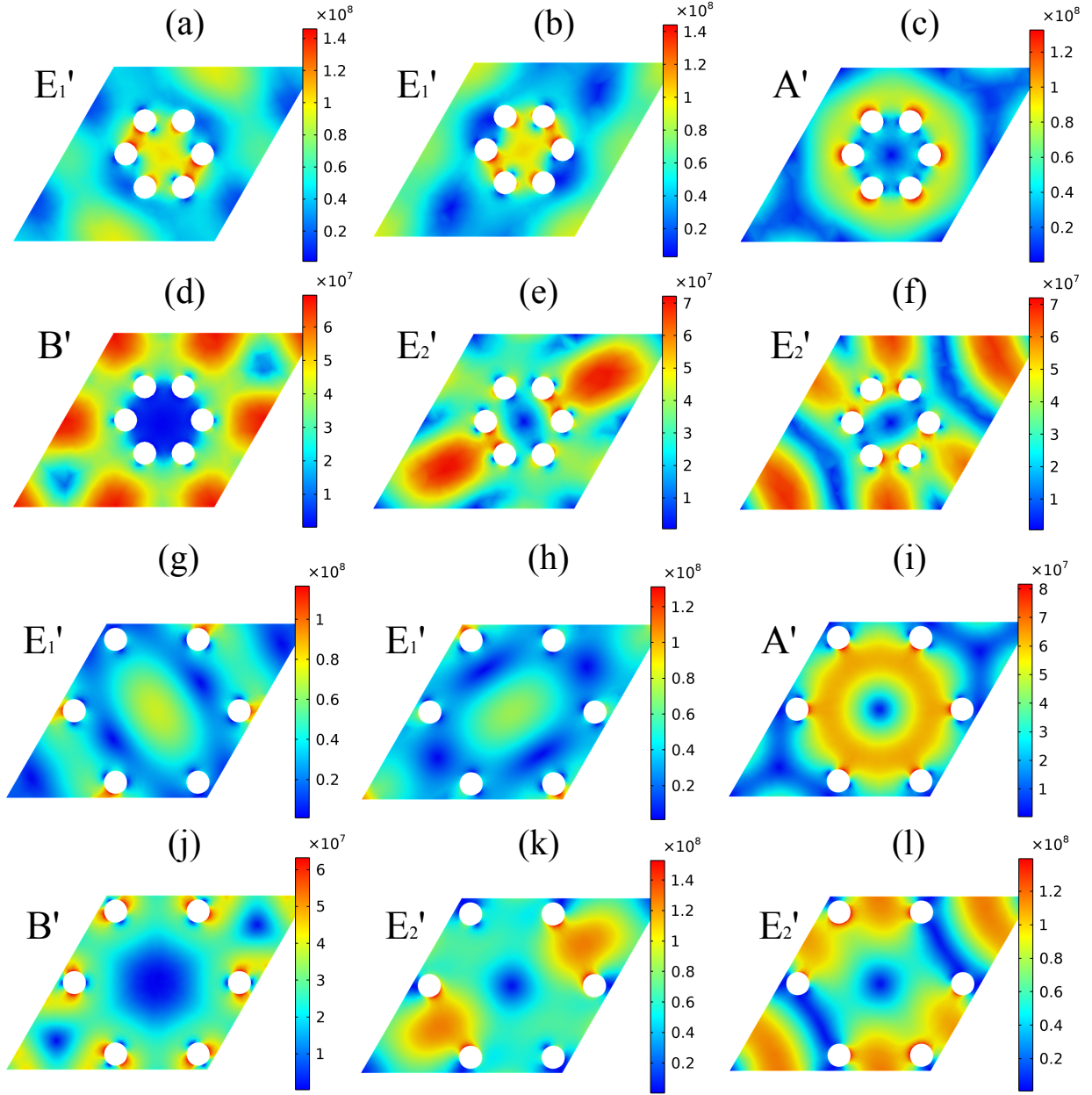

FIG. S2. Electric field magnitudes for the in-plane eigenmodes in the array plane for the triangular lattice of hexamers with (a)-(f) an edge-to-edge distance of 50 nm between particles and with (g)-(l) an edge-to-edge distance of 200 nm between particles. Results were computed with finite element method (COMSOL software) assuming cylindrical gold particles with a diameter of 80 nm and height of 50 nm. Background index of refraction was $n = 1.52$ and the triangular lattice had a periodicity of 680 nm. We indicate the associated irreducible representation in the figures.

In Fig. S2(a-f) we show only the electric field magnitudes (leaving away the arrows that tell directions in Fig. S1) calculated from FEM for the in-plane modes labelled according to the IRs for an edge-to-edge distance $t = 50$ nm. By looking at the modes which have *the largest area of strong intensity fields in the unit cell away from the close vicinity of the nanoparticles*, we see that we could order the modes as $B'$, then the doublets $E'_2$, $E'_1$, and finally $A'$. By comparison with Fig. 2(f) of the main text, we see that this order corresponds exactly to modes with the highest $Q$ factor at $t = 50$ nm ($B'$, $E'_2$, $E'_1$, $A'$). In Fig. S2(g-l) we then show FEM simulations for $t = 200$ nm; we see that the field distribution inside the unit cell changes significantly, despite the symmetry of the mode being the same. This mechanism can explain why the losses of the modes change, and can also contribute to how a given mode can couple to the gain medium, hence the $Q$ factors. Similar reasoning as before would order the modes with the largest area of strong intensity fields away from the nanoparticles as $E'_2$, then $A'$, $E'_1$ and finally $B'$. Again, this order is the same one that is found for the $Q$ factors in Fig. 2(f) of the main text for $t = 200$ nm.

## SAMPLE FABRICATION

The nanoparticle arrays are fabricated on borosilicate microscope slides (75 mm x 25 mm) and a 200 nm thick layer of Poly(methyl methacrylate) (PMMA) is spin coated and baked on the substrate. A 10 nm thick layer of almunium is evaporated on top and the PMMA is patterned with electron beam lithography (EBL). The aluminium is etched in a 1:1 mixture of de-inonized water and AZ351B developer, followed by developing the PMMA in a 3:1 mixture of isopropanol and methyl isobutyl ketone for 15 s. An adhesive layer of titanium (2 nm) is evaporated, followed by 50 nm of gold. A lift-off to remove excess PMMA and metal is done in acetone. For transmission measurements, the samples are immersed in index-matching oil and a coverslide is placed on top. For lasing measurements, IR 140 molecules are dissolved in a 10 mM concentration in a 1:2 mixture of dimethyl sulfoxide and benzyl alcohol, to ensure index-matching to the substrate. The dye solution is injected into a press-to-seal silicone isolator chamber with a thickness of 0.8 mm between the sample slide and a second borosilicate slide. The fabricated arrays have a periodicity of $a = 680$ nm, the height of the nanoparticles is 50 nm and the diameter $d = 80$ nm. The array is in shape of an equilateral triangle with side lengths of $150\mu$m. The edge-to-edge distance $t$ between the particles in the hexamer is varied from $45 - 200$ nm.

## EXPERIMENTAL SETUP

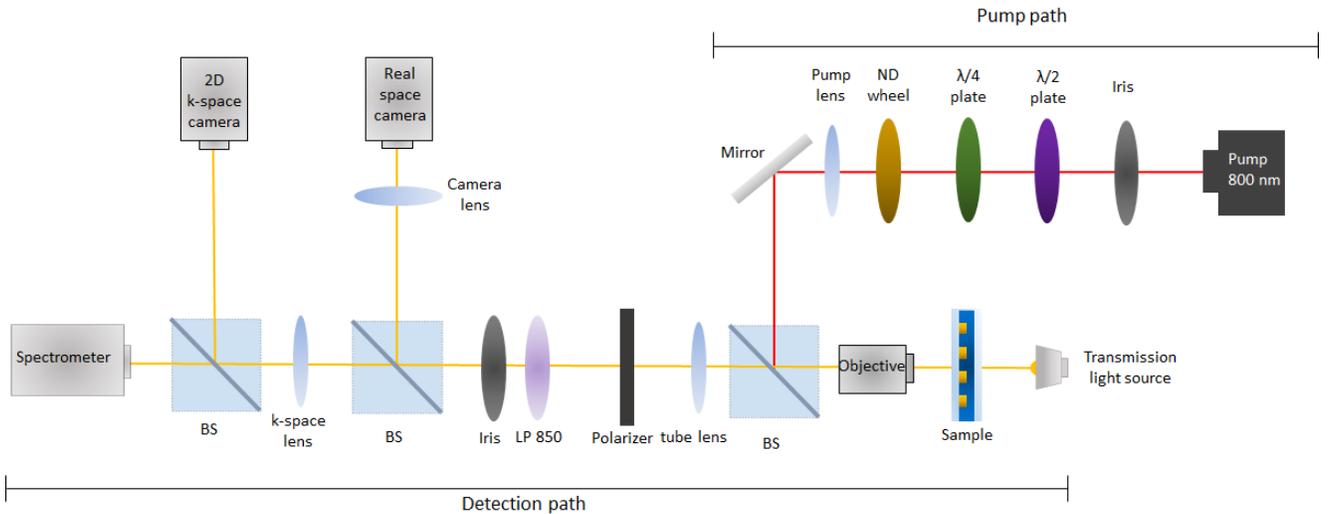

FIG. S3. Experimental setup for lasing and transmission measurements. Angle-resolved spectra, real space images and 2D momentum space images can be simultaneously measured with the two CMOS cameras and the 2D CCD camera in the spectrometer. Here, ND, BS, LP 850 mean neutral density, beam splitter and long pass filter with 850 nm cutoff length, respectively.

The experimental setup is shown in Fig. S3. For lasing measurements, the sample is pumped with a fs laser with 800 nm central wavelength, which is left-circularly polarized. The pump fluence is controlled with a neutral density wheel. For transmission measurements, the sample is illuminated with white light from a halogen lamp.

## LASING EXPERIMENTS: FWHM, LASING ENERGIES, AND SPECTRA

The lasing energies and full width at half max (FWHM) as a function of pump fluence, and spectra at a pump fluence of 0.89 mJ/cm$^2$ are shown in Fig. S4. The energy of the lasing peak ranges from 1.39 eV ($t = 50$ nm), over 1.387 eV ($t = 120$ and 170 nm) to 1.386eV ($t = 200$ nm). Interestingly, there is a switch in the lasing energy for the $t = 200$ nm array at a pump fluence of 0.89 mJ/cm$^2$. The FWHM is between 0.7 and 1.5 meV, only for the $t = 200$ nm array it has a value of 2.3 meV. Looking at the spectra for each array at a pump fluence of 0.89 mJ/cm$^2$ in Fig. S4(c), there are two peaks for the $t = 200$ nm array at 1.386 and 1.388 eV. We attribute the dip in the peak intensity for $t = 200$ nm (see Fig. 2 (b) in the main text) to a mode competition between these two lasing modes. The lower energy peak is identified as the $A'$ mode by the polarization analysis. The energy difference between the two peaks, 2 meV, is the same as the difference between the energies of the $A'$ and $E'_2$ energies in the T-matrix simulations. As $E'_2$ is the highest Q-factor mode, it is plausible that it starts lasing first, while $A'$ can take over due to higher slope efficiency; such mode competition dynamics is common in lasing. In the present case, mode competition is enhanced since the $A'$ and $E'_2$ have high field intensities (see Fig. S2) at complementary spatial positions in the unit cell and can thus utilize gain from different areas.

Note that for $t = 145 - 155$ nm there was no lasing observable. This feature is connected to the array structure: in fact, for $t \approx 146$ nm the array is a perfect honeycomb lattice with a new periodicity of 390 nm. The corresponding Γ-point of this honeycomb lattice is approximately at 2.4 eV, whereas the original Γ-point is at a much lower energy, around 1.4 eV: the Γ-point of the honeycomb structure does not have significant overlap with the gain spectrum of the molecule.

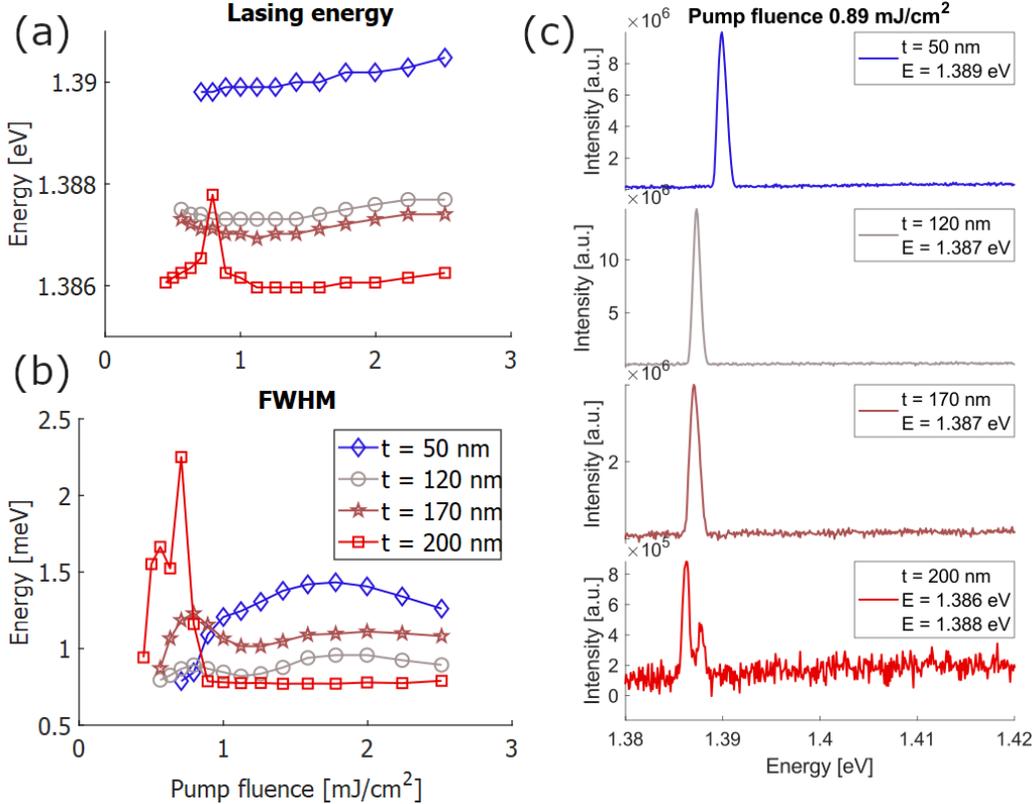

FIG. S4. Lasing in arrays with varying $t$: (a) lasing energy, (b) full width at half max (FWHM), and (c) spectra of the lasing emission at a pump fluence of 0.89 mJ/cm$^2$. Below the lasing threshold we were not able to determine the FWHM due to low intensity, therefore we do not show it in (b).

## COMPLEX ENERGIES OF THE MODES AS A FUNCTION OF THE HEXAMER SIZE

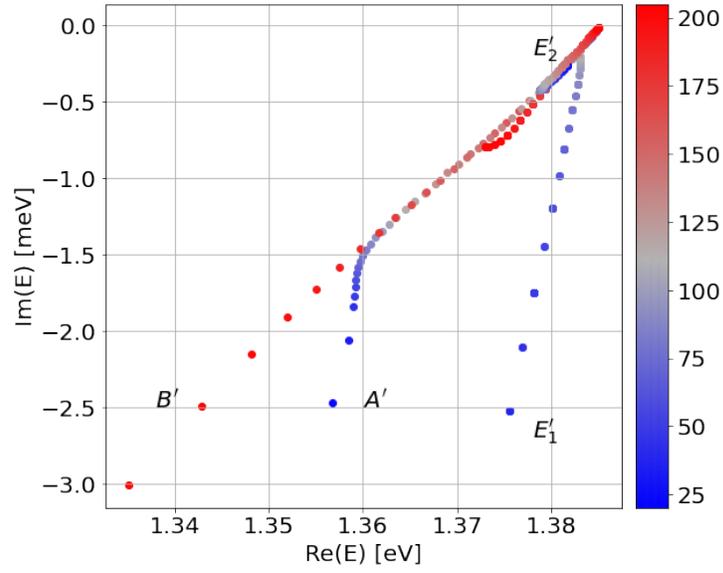

FIG. S5. The modes as in Fig. 2 of the main text but without the energy shifts. The dark modes mostly exist along the same line implying that they have the same loss mechanism. The bright mode, which is known to lose energy by radiating, has a different slope for $t < 125$.

## REAL SPACE IMAGES OF THE ARRAYS IN THE LASING REGIME

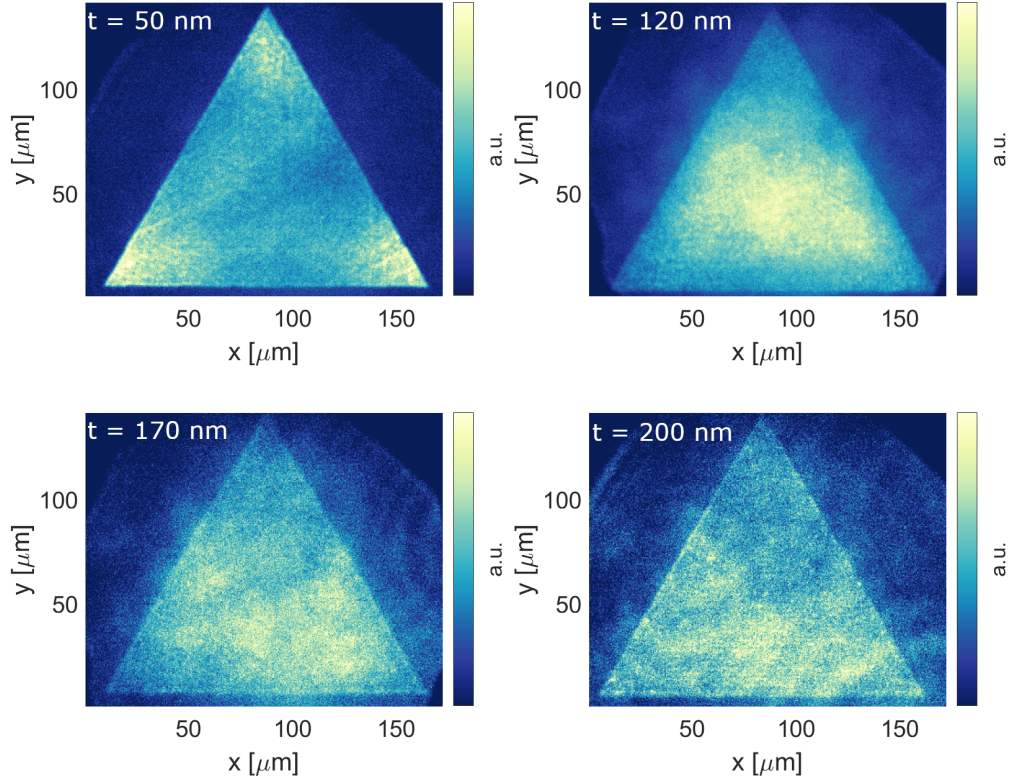

FIG. S6. Real space images of the arrays with different $t$ in the lasing regime. The images were taken at pump fluences for t= 50 nm at 1.41 mJ/cm$^2$, for t = 120 nm at 2.00 mJ/cm$^2$, for t = 170 nm at 2.00 mJ/cm$^2$, and for t = 200 nm at 1.26 mJ/cm$^2$. For t = 50 nm the emission concentrates mostly at the edges of the array, indicating that the edges act as the leakage channel for the quasi-BIC. The bright mode observed for t = 120 nm shows real space emission mainly from the bulk of the array. A bright mode can be observed in the far-field and thus does not require any leakage channel. For t = 170 nm and t = 200 nm the overall emission is quite weak (compare e.g. Fig. 2 (b) and (c) in the main text), and both the edges and the bulk show some emission. In presence of losses, a quasi-BIC can have some emission from the bulk too, as shown by the simulations of [S5].

## TOPOLOGICAL REGIONS OF A QUADRUMER PLASMONIC LATTICE

We have performed T-matrix simulations for the quadrumer plasmonic lattice studied in Ref. [S5], obtaining the energy and the $Q$ factors of the modes as the edge-to-edge distance between nanoparticles in a quadrumer is changed, Fig. S7. Also in this quadrumer case we see a change in the mutual ordering of the modes' $Q$ factors, and a topological transition should occur for edge-to-edge distance $t = 90$ nm. In our previous work [S5] we experimentally studied lasing in quadrumer arrays with a range of different quadrumer sizes as is discussed in the Supplementary Information (see Section "Structural Parameter Dependence"); we did not observe a topological transition. When small quadrumers were considered, we observed lasing at the same lasing energy, with a similar real space emission pattern, from a BIC of topological charge $q = 1$. However when the quadrumer was larger, we did not observe any lasing. This is attributed to a sudden energy change that is visible in Fig. S7; in fact, the energy of the modes around the topological transition quickly becomes not ideal with respect of the gain spectrum of the dye molecules, hence there is not enough gain for lasing in the bright mode with $q = 0$.

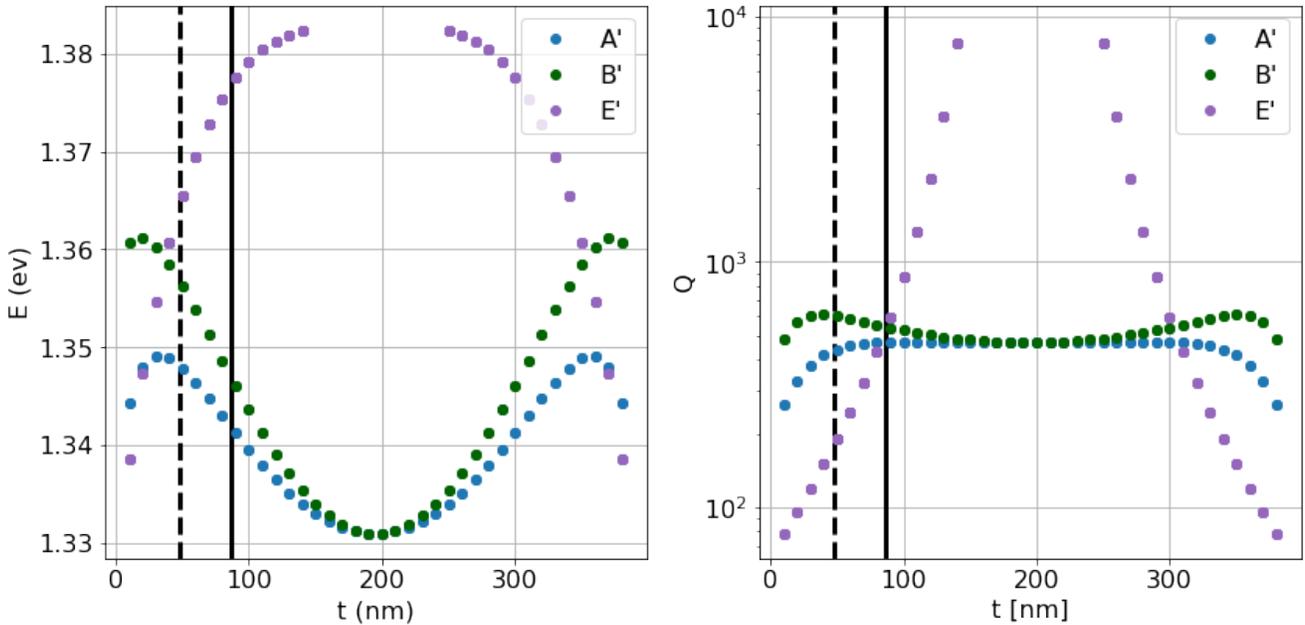

FIG. S7. Modes obtained from the T-Matrix simulations in a quadrumer plasmonic lattice. The dashed line corresponds to the quasi-BIC lasing observed in [S5], and the vertical solid line marks the edge-to-edge distance $t$ of the topological transition.


\end{document}